\documentclass[11pt,twoside]{article}
\usepackage{./asp2014}

\aspSuppressVolSlug
\resetcounters

\begin{document}

\title{Data Reduction of Multi-wavelength Observations}
\author{M.~Pilia,$^1$ A.~Trois,$^1$, A.~P.~Pellizzoni,$^1$ M.~Bachetti,$^1$, G.~Piano,$^2$ A.~Poddighe,$^1$ E.~Egron,$^1$ M.~N. Iacolina,$^1$ A.~Melis,$^1$ R.~Concu$^1$, A.~Possenti,$^1$ \& D.~Perrodin.$^1$}
\affil{$^1$INAF - Osservatorio Astronomico di Cagliari, via della Scienza 5, I-09047 Selargius (Cagliari), Italy; \email{mpilia@oa-cagliari.inaf.it}}
\affil{$^2$INAF - IAPS Roma, Via del Fosso del Cavaliere 100, I-00133 Rome, Italy}

\paperauthor{Maura~Pilia}{mpilia@oa-cagliari.inaf.it}{}{INAF}{Osservatorio Astronomico di Cagliari}{Selargius}{Cagliari}{09047}{Italy}
\paperauthor{Alessio~Trois}{}{}{INAF}{Osservatorio Astronomico di Cagliari}{Selargius}{Cagliari}{09047}{Italy}
\paperauthor{Alberto~Paolo~Pellizzoni}{}{}{INAF}{Osservatorio Astronomico di Cagliari}{Selargius}{Cagliari}{09047}{Italy}
\paperauthor{Matteo~Bachetti}{}{}{INAF}{Osservatorio Astronomico di Cagliari}{Selargius}{Cagliari}{09047}{Italy}
\paperauthor{Giovanni~Piano}{}{}{INAF}{IAPS}{Roma}{Roma}{00133}{Italy}
\paperauthor{Antonio~Poddighe}{}{}{INAF}{Osservatorio Astronomico di Cagliari}{Selargius}{Cagliari}{09047}{Italy}
\paperauthor{Elise~Egron}{}{}{INAF}{Osservatorio Astronomico di Cagliari}{Selargius}{Cagliari}{09047}{Italy}
\paperauthor{Maria~Noemi~Iacolina}{}{}{INAF}{Osservatorio Astronomico di Cagliari}{Selargius}{Cagliari}{09047}{Italy}
\paperauthor{Andrea~Melis}{}{}{INAF}{Osservatorio Astronomico di Cagliari}{Selargius}{Cagliari}{09047}{Italy}
\paperauthor{Raimondo~Concu}{}{}{INAF}{Osservatorio Astronomico di Cagliari}{Selargius}{Cagliari}{09047}{Italy}
\paperauthor{Andrea~Possenti}{}{}{INAF}{Osservatorio Astronomico di Cagliari}{Selargius}{Cagliari}{09047}{Italy}
\paperauthor{Delphine~Perrodin}{}{}{INAF}{Osservatorio Astronomico di Cagliari}{Selargius}{Cagliari}{09047}{Italy}

\begin{abstract}
We are developing a software package that combines gamma-ray data from multiple telescopes with the aim to cross calibrate different instruments, test their data quality, and to allow the confirmation of transient events using different instruments. In particular, we present the first applications using pulsar data from the AGILE and Fermi satellites, and show how we solved the issues relative to combining different datasets and rescaling the parameters of different telescopes. In this way, we extend the energy range observed by a single telescope and can detect fainter objects. As a second step, we apply the pulsar gating technique to the imaging data of the combined satellites in order to look for pulsar wind nebulae. The same procedure is adopted in the radio domain using data from the Sardinia Radio Telescope (SRT). We aim to use similar techniques for multi-frequency datasets spanning a large range of the electromagnetic spectrum. We also present our ongoing work to include an automatic search for gamma-ray counterparts within pulsar search pipelines at radio frequencies.
\end{abstract}

\section{Introduction}
Multi-messenger astronomy is becoming the key to understanding the Universe from a comprehensive perspective. 
In most cases, the data and the technology are already in place, therefore it is important to provide an easily-accessible package that combines datasets from multiple telescopes at different wavelengths.
In order to achieve this, we are working to produce a data analysis pipeline that allows the data reduction from different instruments without needing detailed knowledge of each observation.
Ideally, the specifics of each observation are automatically dealt with, while the necessary information on how to handle the data in each case is provided by a tutorial that is included in the program.

We first focus our project on the study of pulsars and their wind nebulae (PWNe) at radio and gamma-ray frequencies. In this way, we aim to combine time-domain and imaging datasets at two extremes of the electromagnetic spectrum. In addition, the emission has the same non-thermal origin in pulsars at radio and gamma-ray frequencies, and the population of electrons is believed to be the same at these energies in PWNe.
The final goal of the project will be to unveil the properties of these objects by tracking their behaviour using all of the available multi-wavelength data.

\section{Data Reduction}

The first tests were performed using the Crab pulsar. The Crab is the third brightest pulsar in gamma-rays, and is very well studied at all frequencies of the radio domain and outside. The Crab is the youngest known pulsar in our Galaxy, and it is embedded by a PWN with a filled structure called a `plerion'.

For our gamma-ray study, we took advantage of the publicly available archives from the two new generation gamma-ray satellites AGILE \citep{tavani09} and Fermi-LAT \citep{atwood09}.
For our radio study, we reduced data taken in the context of the SRT scientific validation using the new SARDARA backends (SArdinian Roach-2 Digital Architecture for Radio Astronomy, Melis et al. in prep.).

\articlefiguretwo{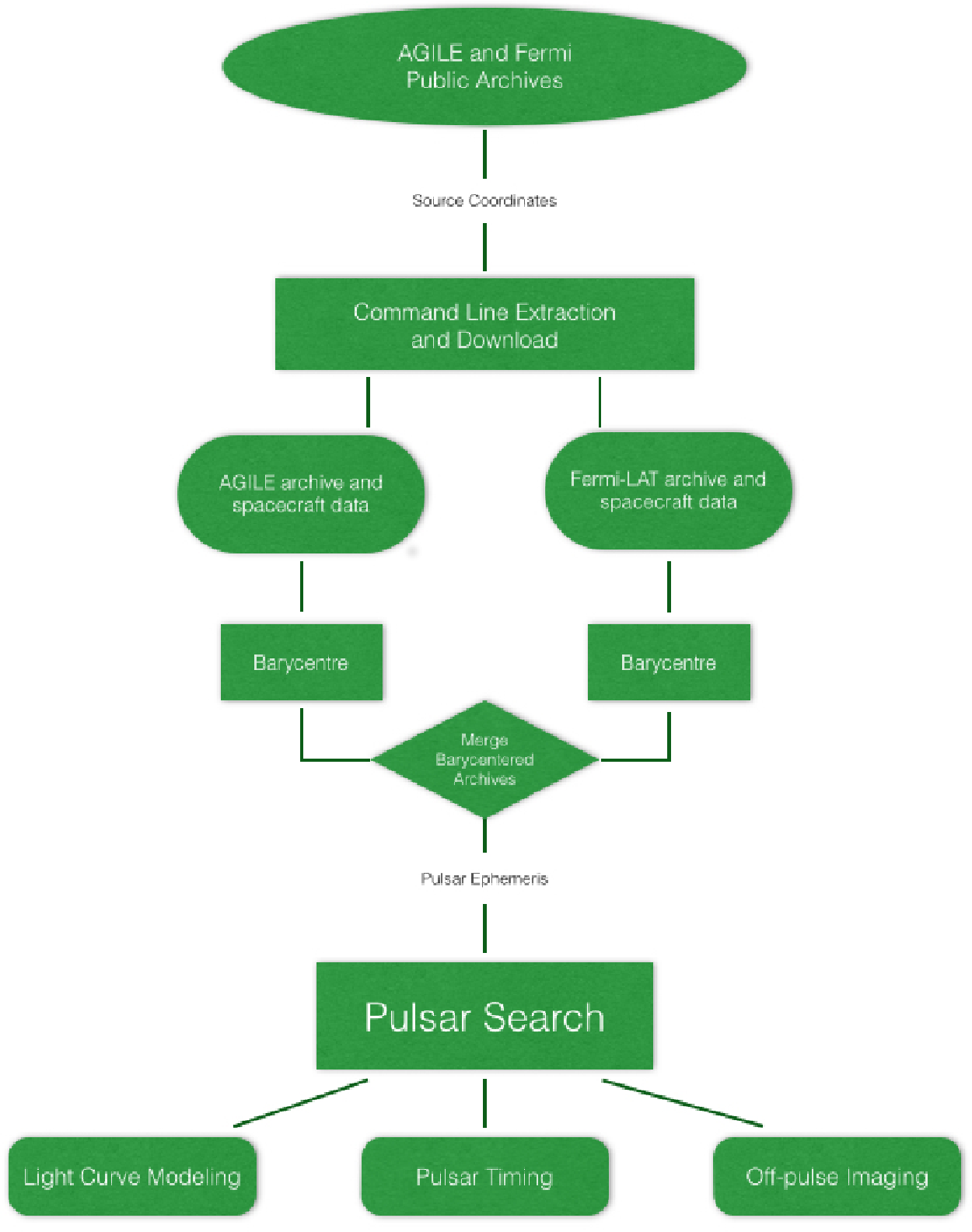}{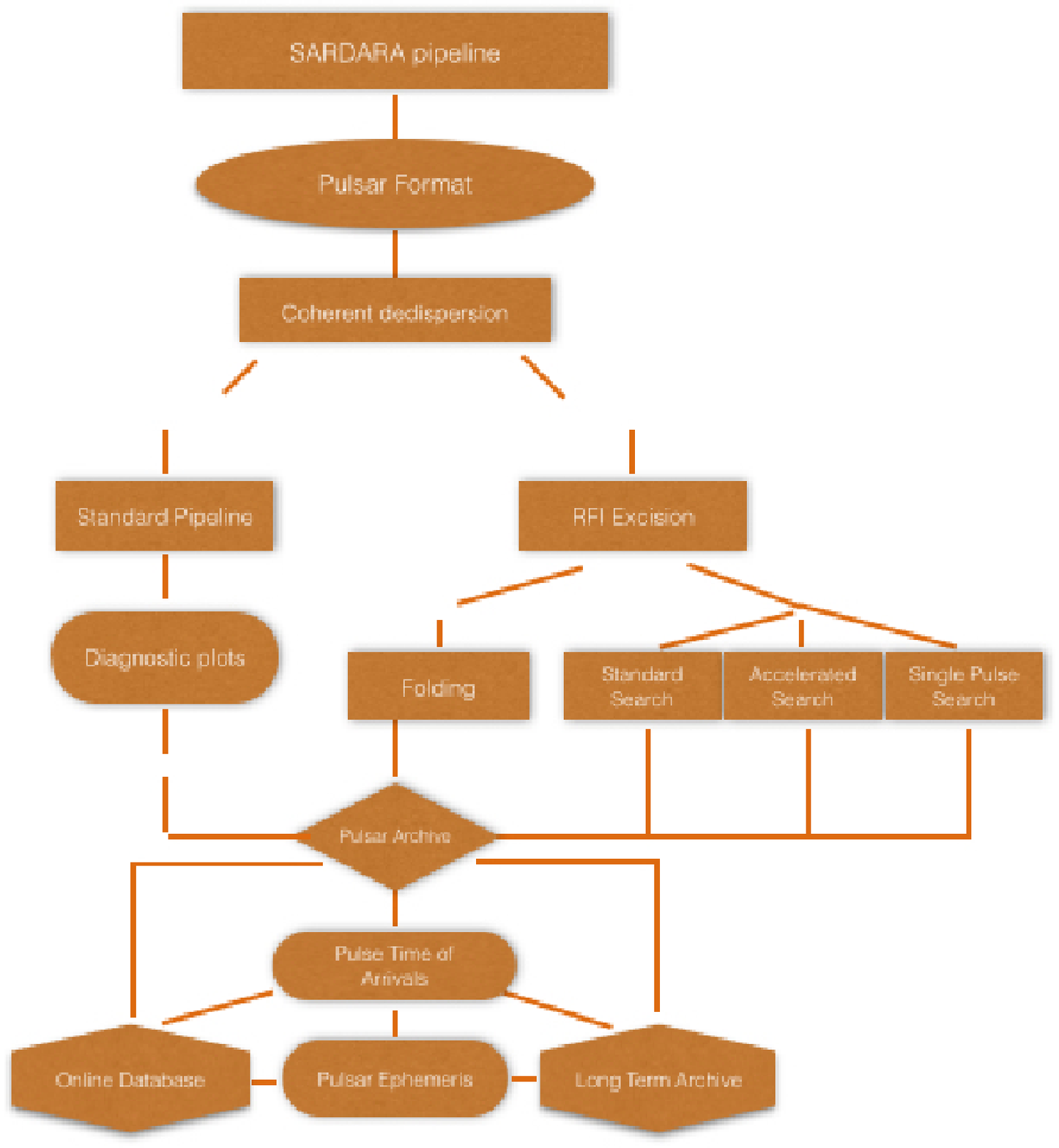}{pipelines}{Schematic views of the pipelines for pulsar analysis.  \emph{Left:} Gamma-rays.  \emph{Right:} Radio.}

\subsection{The pipeline}

We have automated the search of the publicly available gamma-ray data with the creation of a single command line script which, in its simplest approach, takes the coordinates of the source as its input and extracts all of the available observations for an extraction radius of 10 degrees around the source, plus the corresponding spacecraft data.
Given a pulsar ephemeris, we can barycentre the data relative to each satellite and then combine them to do a search in pulse period and period derivative according to a scheme similar to the one described in \citet{pellizzoni09a}.
In the case of a pulsar surrounded by a PWN, we can subtract the pulsed flux from the data and reanalyse the tagged off-pulse data to search for diffuse emission around the source, since the pulsar is usually dominant at medium-energy gamma-rays (see \citealt{pellizzoni10}).

In order to regularly follow pulsars at all frequencies, it is fundamental to keep up-to-date pulsar ephemerides. Nowadays, this is possible using gamma-ray data ({see \citealt{kerr15}); traditionally, when possible, an accurate ephemeris is obtained by the regular timing of pulsars using a radio telescope.
We aim to create a pulsar ephemeris database that can be used for pulsar analysis, both with SRT and gamma-ray data.

The radio observations that are performed locally are digitised using a ROACH-2 backend\footnote{https://casper.berkeley.edu/wiki/ROACH-2\_Revision\_2} (SARDARA) and then converted into the {\tt psrfits} format \citep{hotan04}, which is easily readable from all standard pulsar analysis packages. 
Radio frequency interference is mitigated in the data, which are subsequently dedispersed and folded.
For known pulsars, we will produce integrated pulse profiles and compute times of arrival by comparison with a standard template. Over time, this will allow us to obtain a timing model and produce an ephemeris.

The purpose of our project is also to find new pulsars from the error boxes of the gamma-ray-only pulsars or unidentified gamma-ray sources. For this purpose, we are developing a search pipeline by integrating the already available programs from the {\tt presto} \citep{ransom01} and {\tt psrchive} \citep{vanstraten12} suites for pulsar analysis.
We are first completing a performance study of SRT to estimate its flux limits. This is also a first fundamental step for future searches of fast transients with SRT.

\subsection{First Results}
Here we present the first results obtained from our data collection and analysis.

The left-hand side of Figure \ref{results} shows the Crab pulsar light-curve at $E > 100$\,MeV that was obtained from the combination of 1\,yr of AGILE and Fermi-LAT data using a public ephemeris from the Fermi database. While this is only a test, much better results can be obtained by combining a longer data span, for which a precise ephemeris covering the whole time interval is needed.

On the right-hand side of Figure \ref{results}, we present the first image of a PWN observed with SRT. It was obtained at 6.5 GHz, with 1\,GHz of bandwidth in a 20\,-min observation using the SARDARA backend. The image can be refined using pulsar gating techniques. This technique will also be applied to look for pulsars in imaging data, especially at lower radio frequencies.

\articlefiguretwo{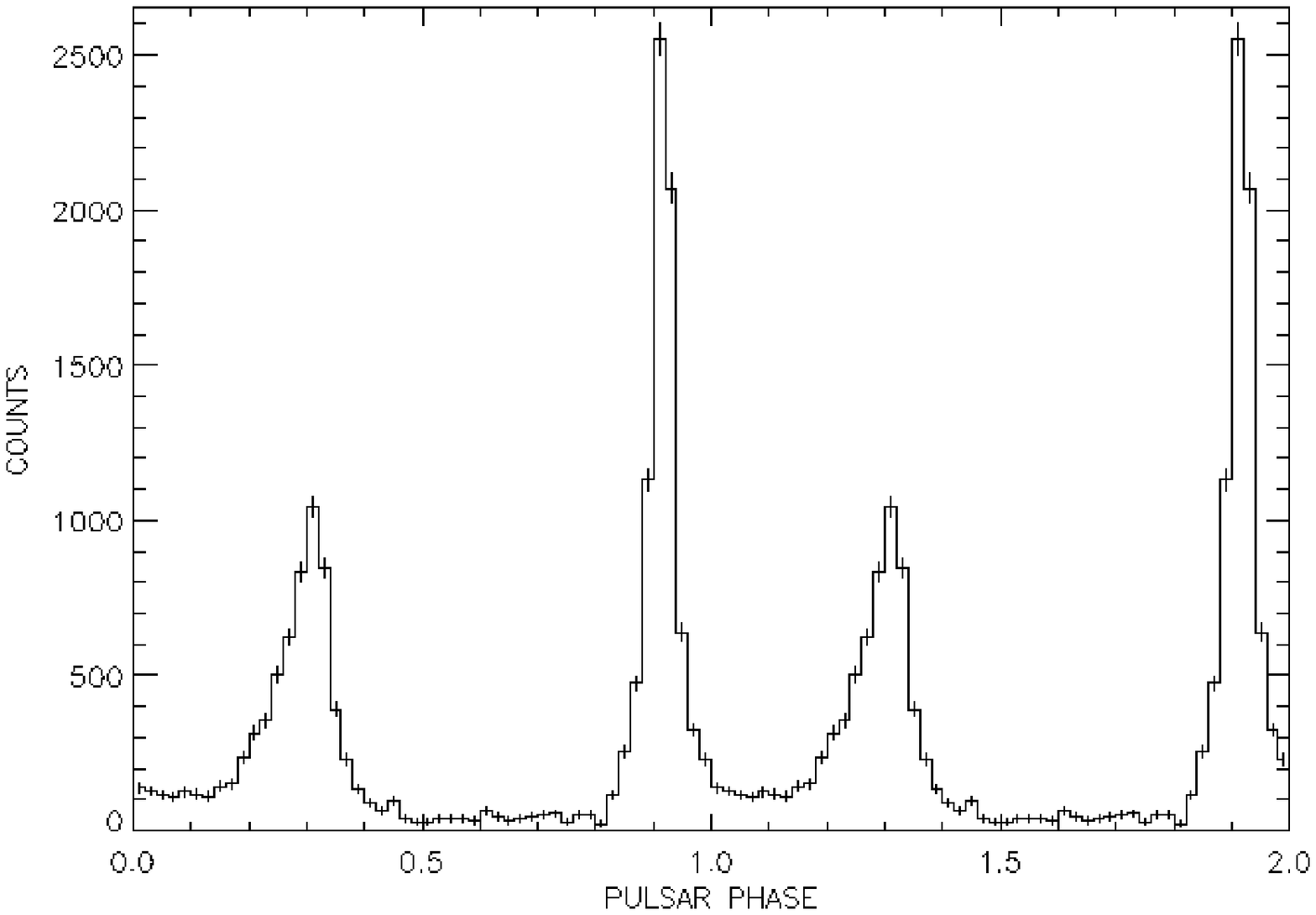}{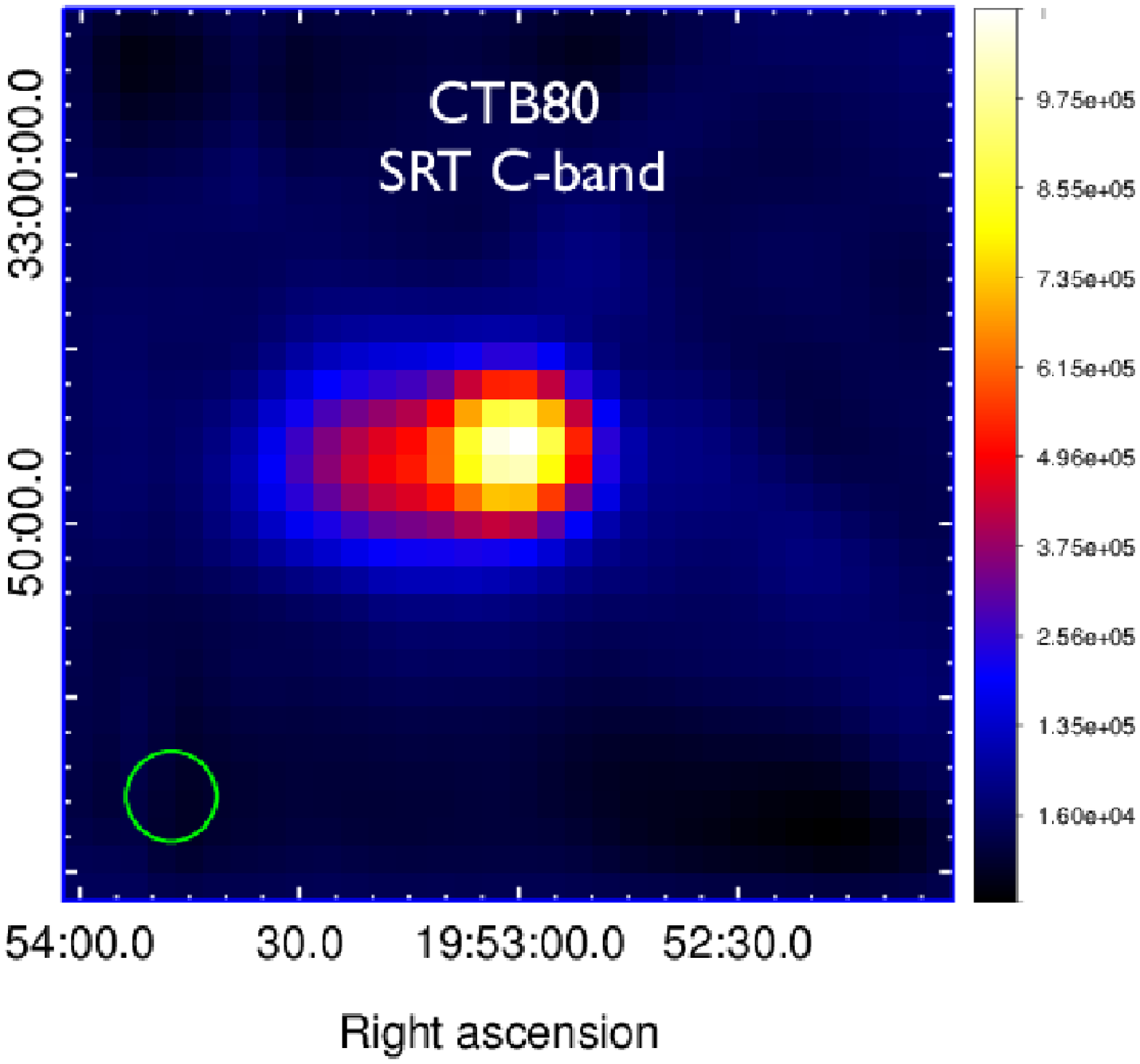}{results}{First results of our pipelines. \emph{Left:} Crab pulsar light-curve at $E > 100$\,MeV from combined AGILE and Fermi data.  \emph{Right:} 6.5 GHz imaging of PWN CTB~80 with SRT.}

\section{Conclusions and Future Work}

Given a pulsar ephemeris, we are already able to search the public archives of AGILE and Fermi and to obtain pulsar light curves from each instrument. We have managed to combine in phase both datasets so as to cross check the long term performance stability of both instruments, obtain a larger dataset and, finally, improve our sensitivity to structures at sub-millisecond level in the pulsar profiles.
Gamma-ray observations with declinations above $\sim -30$ degrees can be followed up with radio observations, and vice-versa: future radio searches for pulsars and fast transients with SRT will be automatically cross-searched in gamma-rays for counterparts thanks to an embedded database in the analysis pipeline.
From all of the available data, we collect time-stamps in order to create timing models for the pulsars and make them available for subsequent analyses.
Once we obtain a good solution for a pulsar, we can apply the technique of subtraction of the pulsed flux to look for diffuse emission around the pulsar: it is typically a PWN, but pulsar `tails' have also been observed in X-rays.

With these procedures, we aim to considerably reduce the time and manpower needed to perform pulsar surveys and follow-up studies of observations at different wave-bands.
As a future development, we envisage the extension of the number of wave-bands involved in our combined databases, starting from the X-rays, which already provide the broadest set of publicly available data products, and subsequently millimeter and microwaves.
The novelty of our approach relies not in the planned availability of data products (i.e. pulsar ephemerides and light curves or images from standard automatic pipelines run on the data), but in the possibility to retrieve raw data at different wavelengths and analyse them, all through a procedure that will be mostly wavelength-independent.

\acknowledgements The authors acknowledge support from the Autonomous Region of Sardinia (RAS) for the project "Development of a Software Tool for the Study of Pulsars from Radio to Gamma-rays using Multi-mission Data" CRP-25476. The SARDARA backend is financed by RAS through the Tender-7 funding programme. This work is partially based on commissioning observations  with SRT operated
by INAF. For observations at SRT, we also credit the Astrophysical Validation Team\footnote{http://www.srt.inaf.it/astronomers/astrophysical-validation-team/}.

\def\pasa {PASA}
\bibliography{biblio}  

\end{document}